\documentclass[twocolumn,pra,aps,showpacs,showkeys,superscriptaddress,10pt]{revtex4-1} 
\usepackage{amsmath,amssymb,graphicx}
\usepackage{hyperref}
\usepackage{epsfig}
\begin{document}
\title{\normalsize Control of Fano resonances and slow light using Bose-Einstein condensates in a nanocavity}
\author{M. Javed Akram}
\email{mjakram@qau.edu.pk}
\affiliation{Department of Electronics, Quaid-i-Azam University, 45320 Islamabad, Pakistan}
\author{Fazal Ghafoor}
\email{rishteen@yahoo.com}
\affiliation{Department of Physics, COMSATS Institute of Information Technology, 45550 Islamabad, Pakistan}
\author{M. Miskeen Khan}
\email{mmiskeenk16@hotmail.com}
\affiliation{Department of Electronics, Quaid-i-Azam University, 45320 Islamabad, Pakistan}
\author{Farhan Saif}
\email{farhan.saif@fulbrightmail.org}
\affiliation{Department of Physics, Quaid-i-Azam University, 45320 Islamabad, Pakistan}
\begin{abstract}
In this study, a standing wave in an optical nanocavity with Bose-Einstein condensate (BEC) constitutes a one-dimensional optical lattice potential in the presence of a finite two bodies atomic interaction. We report that the interaction of a BEC with a standing field in an optical cavity coherently evolves to exhibit Fano resonances in the output field at the probe frequency. The behaviour of the reported resonance shows an excellent compatibility with the original formulation of asymmetric resonance as discovered by Fano [U. Fano, \href{http://journals.aps.org/pr/abstract/10.1103/PhysRev.124.1866}{Phys. Rev. 124, 1866 (1961)}]. Based on our analytical and numerical results, we find that the Fano resonances and subsequently electromagnetically induced transparency of the probe pulse can be controlled through the intensity of the cavity standing wave field and the strength of the atom-atom interaction in the BEC. In addition, enhancement of the slow light effect by the strength of the atom-atom interaction and its robustness against the condensate fluctuations are realizable using presently available technology.
\end{abstract}
\maketitle 
\section{Introduction}
\enlargethispage{7\baselineskip}
Fano resonance was discovered as the asymmetric feature of a photoionization cross section in an atom. It is attributed to the destructive interference between the probability amplitudes of direct photoionization, and through the auto-ionizing-state indirect photoionization to the ionizing continuum \cite{Fano,Kivshar,GSAgarwal,Scully,Ott}. Since its discovery, the asymmetric Fano resonance has been a characteristic feature of interacting quantum systems, such as quantum dots \cite{Johnson,Bird}, plasmonic nanoparticles \cite{Chong}, photonic crystals \cite{Rybin,Christ}, phonon transport \cite{Tang}, Mach-Zhender-Fano interferometry \cite{Mir,Kivshar}, whispering-gallery-modes \cite{Lei,Peng}, extreme ultraviolet (XUV) attosecond spectroscopy \cite{Liao}, electromagnetic metamaterials \cite{Wallauer} and bio-sensors \cite{Yanik,Ameen}. Fano resonances are characterized by a steeper dispersion than conventional Lorentzian resonances \cite{Chong,Kivshar}, which make them promising for local refractive index sensing applications \cite{Wallauer}, to confine light more efficiently \cite{Kivshar} and for surface enhanced Raman scattering (SERS) \cite{Ye}. Besides these applications, Fano resonances have also been used for enhancing the biosensing performance \cite{Wu,Verellen}, enhanced light transmission \cite{Zhou}, slow light \cite{Zhang}, and classical analog of electromagnetically induced absorption (EIA) \cite{Taubert}. More recently, Heeg et al. \cite{Heeg} reported the use of Fano resonances for interferometric phase detection and x-ray quantum state tomography which provide new avenues for structure determination and precision metrology \cite{Heeg}.

In parallel, cavity optomechanics \cite{AKM,Meystre} has cemented its place in the present-day photonic technology \cite{Paternostro,Rabl}; and serves as basic building block from quantum state-engineering to the quantum communication networks \cite{Ladd,Cirac,Paternostro2}. Due to the ubiquitous nature of the mechanical motion, such resonators couple with many kinds of quantum devices, that range from atomic systems to the solid-state electronic circuits \cite{Zoller2,Tian,Akramsr,Milburn}. More recently, a new development has been made in the field of levitated optomechanics \cite{Chang2,Monteiro,Millen}, in which the mechanical oscillator is supported only by the light field. These platforms offer the possibility of a generation of highly-sensitive sensors, which are able to detect (for example) weak forces, with a precision limited only by quantum uncertainties \cite{Millen}. Fano resonances \cite{Qu,Akram}, optomechanically induced transparency \cite{Weis} with single \cite{Agarwal} and multiple windows \cite{Akram,Nori,Ghafoor2}, superluminal and subluminal effects \cite{Hau,Lipson,Kash,Scully2,AH,Akram2,Stenner,Tarhan,Zhu,Zhan,Jiang,Hakuta,Segard,Menon} have also been observed in optomechanical systems, where nano dimensions and normal environmental conditions have paved the new avenues towards state-of-the-art potential applications, such as imaging and cloaking, telecommunication, interferometry, quantum-optomechanical memory and classical signal processing applications \cite{Chang,Boyd3,Milonni,Ghafoor1}.

Merging optomechanics with cold atomic systems \cite{Hunger}, for instance, Bose-Einstein condensates \cite{Cornell,Aspect,Carr,Chen,Jing,Steinke,Esslinger,Chiara,Bose,Asjad,Murch,Markku} and degenerate cold atom Fermi gases \cite{Fermions} leads to hybrid optomechanical systems. Transition from Mott insulator state to superfluidity of atoms \cite{Creffield} in an optical lattice coupled to a vibrating mirror has been analyzed, as an example of a strongly interacting quantum system subject to the optomechanical interaction \cite{Larson}. Recently, a comprehensive strategy for using quantum computers to solve models of strongly correlated electrons, using the Hubbard model as a prototypical example has been reported \cite{Wecker}. More recently, interferometric phase detection controlled by Fano resonances and manipulation of slow light propagation have been reported in the x-ray regime \cite{Heeg,Heeg2}. Owing to the significant importance of Fano resonances and slow light in the control of transmission and scattering properties of electromagnetic waves in nano scale devices, we explain the control of the asymmetric sharp and narrow resonances in optomechanics with BEC. In the present paper: (i) we report the emergence of Fano resonances in the presence of BEC loaded in a nano cavity, where standing cavity field forms the one-dimensional lattice potential \cite{Ritsch}. (ii) Upon tuning the resonances versus a wide range of system parameters, we find that the behavior of the resonance reveals an excellent compatibility with the original formulation of asymmetric resonance as discovered by Ugo Fano \cite{Fano}. (iii) Moreover, we discuss subluminal behavior of the probe field in the system, and explain its parametric dependence. (iv) Unlike previous schemes, we note that the magnitude of slow light can be enhanced by continuously increasing the atom-atom interaction, as well as it is less affected by the condensate fluctuations. This reflects the advantage of present scheme over earlier schemes \cite{AH,Akram2,Tarhan,Zhu,Zhan,Jiang}, which may help to realize longer optical storage (memory) applications \cite{Simon,Jobez,Phillips}.

The rest of the paper is organized as follows: In Sec.~\ref{sec2}, we present the system and formulate the analytical results to explain the Fano resonances and slow light effect based on standard input-output theory. Section~\ref{sec3} is devoted to compare obtained analytical results with numerical results, where emergence of the Fano resonances is demonstrated. In Sec.~\ref{sec4}, we explain slow light and its enhancement in the probe transmission. Finally in Sec.~\ref{sec5}, we conclude our work.
\section{The Model Formulation}\label{sec2}
\begin{figure}[b]
\includegraphics[width=0.45\textwidth]{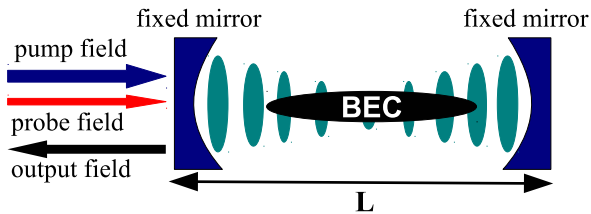}
\caption{(Color online) The schematic representation of the system: an optomechanical system with BEC confined in an optical cavity with fixed mirrors. A strong driving field of frequency $\omega_l$ and a weak probe field of frequency $\omega_p$ are simultaneously injected into the cavity.} \label{model}
\end{figure}
We consider an optomechanical system (OMS) with an elongated cigar-shaped Bose-Einstein condensate (BEC) of $N$ two-level ultracold $^{87}Rb$ atoms (in the $|F = 1\rangle$ state) with $m$ being mass of single atom and $\omega_a$ the transition frequency $|F = 1\rangle \rightarrow |F' = 2\rangle$ of the $D_2$ line of $^{87}Rb$. The nano optical resonator is composed of fixed mirrors which contain a standing wave with frequency $\omega_c$. Hence, we find one dimensional optical lattice potential along the cavity axis, which is coupled strongly with the BEC (see Fig.~\ref{model}). The system is coherently driven by a strong pump-field and a weak probe-field of frequencies $\omega_l$ and $\omega_p$, respectively. The optomechanical-Bose-Hubbard (OMBH) Hamiltonian of the system can be written as \cite{Bhattacherjee,Bhattacherjee2,Ritsch},
\begin{eqnarray}
&H_T & = \hbar \Delta_c c^{\dagger}c  + E_0 \sum_i b_i^\dagger b_i + J_0(\hbar U_0 c^\dagger c + V_{cl})\sum_i b_i^\dagger b_i \nonumber \\
&+& \frac{U}{2}\sum_i b_i^\dagger b_i^\dagger b_i b_i + i\hbar \Omega_l (c^\dagger - c) + i\hbar\varepsilon_p(e^{-i\delta t}c^\dagger - e^{i\delta t}c) \nonumber
\end{eqnarray}
where, the first term represents the free Hamiltonian of the single cavity field mode with the creation (annihilation) operator $c^\dagger$ ($c$). The second term describes the on-site kinetic energy of the condensate with the creation (annihilation) operators $b_i^{\dagger}$ ($b_i$) at the $i$th site. The third term shows the interaction of the condensate with the cavity field. Here, the parameter $U_{0}=\frac{g_{0}^{2}}{\Delta_{a}}$ illustrates the optical lattice barrier height per photon and represents the atomic back-action on the field, $g_0$ is the atom-field coupling and $V_{cl}$ is the classical potential \cite{Ritsch}. 
The fourth term describes the two-body atom-atom interaction, where $U$ is the effective on-site atom-atom interaction energy. Finally, the fifth and the sixth terms account for the intense pump laser field and the weak probe laser field, respectively. Here, $\Omega_l=\sqrt{2\kappa P_{l}/ \hbar \omega_l}$ and $\varepsilon_p = \sqrt{2\kappa P_{p}/ \hbar \omega_p}$ are amplitudes of the pump and probe fields, respectively, where $P_l$ ($P_p$) is the power of the pump (probe) field, and $\kappa$ is the decay rate of the cavity field. Moreover, $\Delta_c=\omega_c-\omega_l$ and $\delta=\omega_p-\omega_l$ are the respective detuning of the cavity field and the probe field, with pump field frequency $\omega_l$, respectively.
Moreover, 
\begin{eqnarray}
& E_0 =\int d^3x w(\vec{r}-\vec{r}_i){(\frac{-\hbar \nabla^2}{2m})}w(\vec{r}-\vec{r}_i), \notag\\
& J_0=\int d^3x w(\vec{r}-\vec{r}_i)\cos^2(kx)w(\vec{r}-\vec{r}_i), \notag\\
& U = \frac{4\pi a_s\hbar^2}{m} \int d^3x w|\vec{r}|^4, \label{H2}
\end{eqnarray}
where in Eq.~(\ref{H2}), $E_0$ and $J_0$ describes the effective on-site energies of the condensate, defined in terms of the condensate atomic Wannier functions $w(\vec{r}-\vec{r}_i)$, where $k$ is the wave vector, and $a_s$ is the two-body $s$-wave scattering length \cite{Ritsch}. For a detailed analysis of the system, we include photon losses in the system and decay rate associated with the condensate mode, $\kappa$ and $\gamma_b$, respectively. Thus, the dynamics of the system can be described by the following quantum Langevin equations:
\begin{eqnarray}
&& \dot c = -(\kappa + i \Delta_{c})c - iU_{o} J_{o} c\sum_{i} b^{\dagger}_{i} b_{i} + \Omega_l + \varepsilon_p e^{-i\delta t} + \sqrt{2 \kappa}c_{in}, \nonumber \\
&& \dot b_{i}=-\frac{iE_{o}}{\hbar} b_{i} - iJ_{o}\left[ \frac{V_{cl}}{\hbar}+U_{o}c^{\dagger} c \right]b_{i}-i\frac{U}{\hbar}b^{\dagger}_{i} b_{i} b_{i}- \gamma_{b} b_{i} + \sqrt{2 \gamma_b} b_{in}, \nonumber \label{m22}
\end{eqnarray}
where, $c_{in}$ and $b_{in}$ are the input noise operators associated with the input field and the condensate mode, respectively. To linearize the above set of equations, we write each canonical operator of the system as a sum
of its steady state mean value and a small fluctuation as $c = c_s + \delta c$ and $b = b_s + \delta b$. We define the quadratures of the mechanical mode of the condensate by defining the Hermitian operators, that is, $q = (\delta b + \delta b^{\dagger})/\sqrt{2}$ and $p = (\delta b - \delta b^{\dagger})/i \sqrt{2}$. Moreover, here we assume the negligible tunneling, and
hence we drop the site index $i$ from the bosonic operators \cite{Bhattacherjee,Ritsch}. Thus, the linearized set of quantum Langevin equations is:

\begin{eqnarray}
&& \ddot{q} + \gamma_b \dot{q} + \omega^2_b q= - g(U_{eff}+\nu)(\delta c + \delta c^{\dagger}) +\xi_{in}, \nonumber\\
&& \delta \dot{c}= - (\kappa + i\Delta) \delta c - i gq + \Omega_l + \varepsilon_p e^{-i\delta t}+\sqrt{\kappa}\delta c_{in}, \label{m2}
\end{eqnarray}

where $\Delta=\Delta_c - U_{0}NJ_{0}$ is the effective detuning of the cavity field, $g=2U_{0}J_{0}\sqrt{N}|c_s|^2$, $\omega_b=\sqrt{(\nu + U_{eff})(\nu + 3 U_{eff})}$, $U_{eff}=\frac{UN}{\hbar M}$, $\nu=U_0 J_0 |c_s|^2+\frac{V_{cl}J_0}{\hbar}+\frac{E_0}{\hbar}$, $\xi_{in}=(b_{in}+b_{in}^\dagger)/2$, and $N$ represents the total number of atoms in $M$ sites. In order to study Fano resonances and slow light, here we are interested in the mean response of the coupled system to the probe field in the presence of the pump field, we do not include quantum fluctuations which are averaged to zero \cite{Agarwal,Akram2}. This is similar to what has been treated in the context of EIT where one uses atomic mean value equations, and all quantum fluctuations due to both spontaneous emission and collisions are neglected \cite{Agarwal,Akram}. In order to obtain the steady-state solutions of the above equations, we make the ansatz \cite{Boyd}:
\begin{eqnarray}
&&\langle \delta c \rangle = c_- e^{-i\delta t} + c_+ e^{i\delta t},  \notag \\
&&\langle q \rangle= q_- e^{-i\delta t} + q_+ e^{i\delta t},\label{cmean}
\end{eqnarray}
where $c_\pm$ and $q_\pm$ are much smaller than $c_s$ and $q_s$ respectively, and are of the same order as $\varepsilon_p$. By substituting Eq.~(\ref{cmean}) into Eqs.~(\ref{m2}), respectively, and taking the lowest order in $\varepsilon_p$ but all orders in $\Omega_l$, we get
\begin{equation}\label{cs}
c_s=\dfrac{\Omega_l}{\kappa + i\Delta},
\end{equation}\label{cm}
\begin{equation}
c_-=\frac{[\kappa + i (\Delta - \delta)](\delta^2 - i \delta\gamma_{b}-\omega^2_{b}) + ig(\nu + U_{eff})}{[\kappa^2 + \Delta^2 - \delta(\delta+i\kappa)][\delta^2 - i \delta\gamma_{b}-\omega^2_{b}]+2\Delta g (\nu + U_{eff})}, \label{eq6}
\end{equation}
In order to study the optical properties of the output field, we use the standard input-output relation viz. \cite{Weis}, $c_{out}(t) = c_{in}(t) - \sqrt{2\kappa}c(t)$. Here, $c_{in}$ and $c_{out}$ are the input and output operators, respectively. We can now obtain the expectation value of the output field as,
\begin{equation}
\langle c_{out}(t) \rangle = (\Omega_l-\sqrt{2\kappa} c_s) + (\varepsilon_p - \sqrt{2\kappa} c_-) e^{-i\delta t} - \sqrt{2\kappa} c_+ e^{i\delta t}.
\end{equation}
Note that, in analogy with Eq.~(\ref{cmean}), the second term (on right-hand-side) in the above expression corresponds to the output field at probe frequency $\omega_p$ via the detuning $\delta=\omega_p - \omega_l$. Hence, the real and imaginary parts of the amplitude of this term accounts for absorption and dispersion of the whole system to the probe field. Moreover, the transmission of the probe field, which is the ratio of the returned probe field from the coupling system divided by the sent probe field \cite{AH}, can be obtained as
\begin{equation}
t_p(\omega_p) = \frac{\varepsilon_p - \sqrt{2\kappa} c_-}{\varepsilon_{p}} = 1- \frac{\sqrt{2\kappa} c_-}{\varepsilon_p}.
\end{equation}
For an optomechanical system, in the region of the narrow transparency window, the propagation dynamics of a probe pulse sent to the coupled system greatly alters due to the variation of the complex phase picked up by its different frequency components. The rapid phase dispersion, that is, $\phi_t(\omega_p) = arg[t_p(\omega_p)]$, can cause the transmission group delay given by \cite{AH,AKM,Akram2}:
\begin{equation}
\tau_g = \dfrac{d\phi_t(\omega_p)}{d\omega_p}=\dfrac{d\{arg[t_p(\omega_p)]\}}{d\omega_p}. \label{phase}
\end{equation}
\section{Fano Resonances in the output field}\label{sec3}
In this section, the phenomenon of the asymmetric Fano resonances is explained.
\begin{figure*}[ht]
\includegraphics[width=0.4\textwidth]{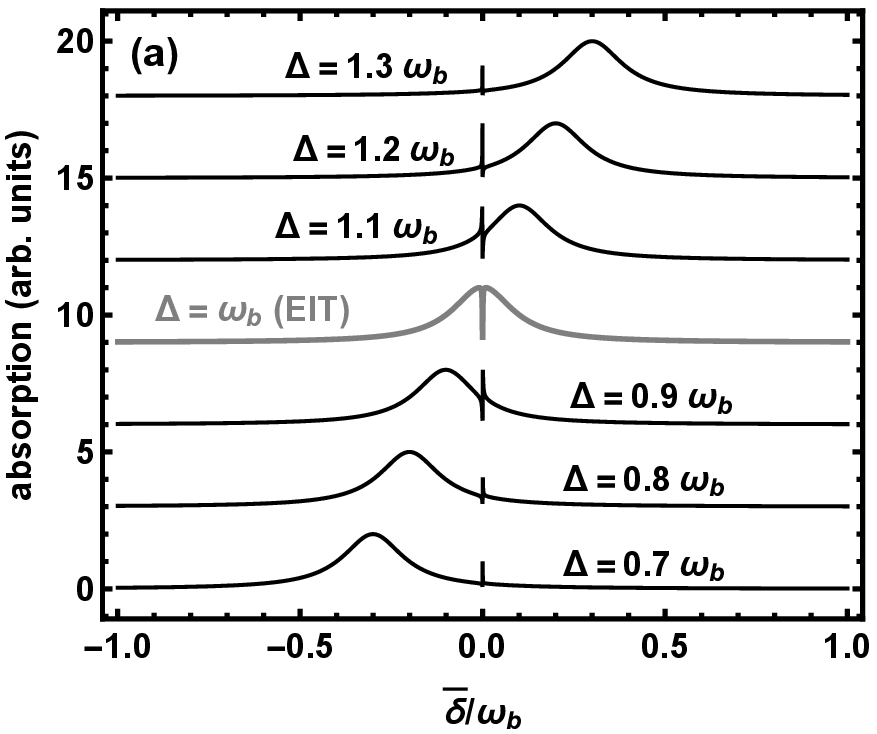}
\includegraphics[width=0.4\textwidth]{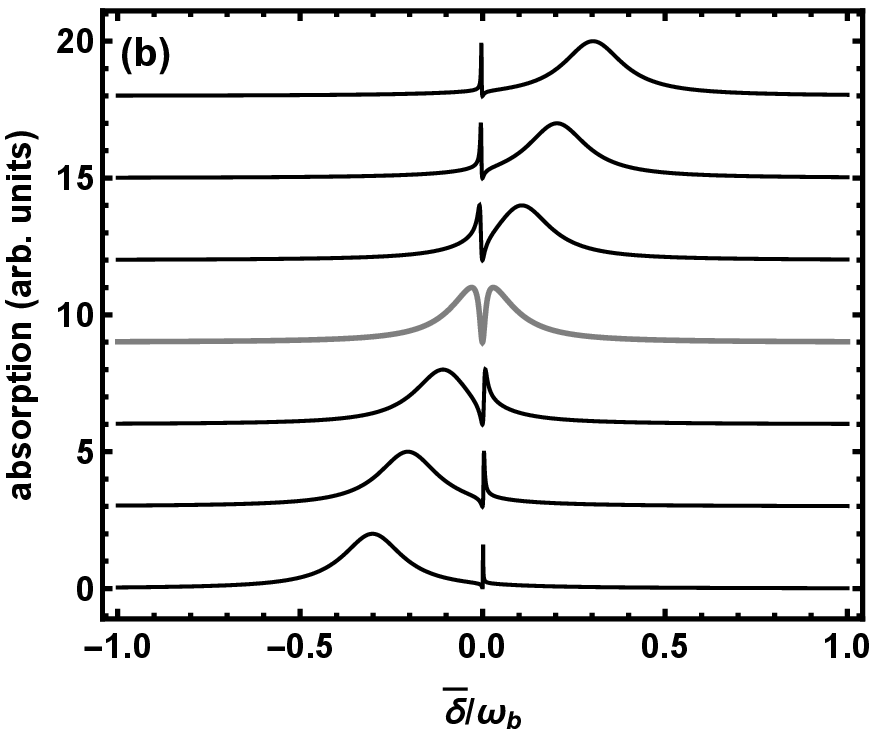}
\caption{(Color online) Fano interference and EIT for (a) $g=0.1\omega_b$ and (b) $g=1\omega_b$: Spectral tuning of the optical response of the system from asymmetric Fano resonance to EIT. When the effective cavity detuning in the BEC-cavity setup is non-resonant with the effective frequency $\omega_b$ i.e. $\Delta \neq \omega_b$, the transmission exhibits asymmetric Fano resonances. At resonant detuning ($\Delta=\omega_b$), a transparency (EIT) window appears. The rest of the parameters are \cite{Esslinger,Murch}: $\kappa= 0.1\omega_b$, $U_{eff}=\omega_b$, $\gamma_b/2\pi=7.5\times 10^{-3}$ Hz, $\nu/2\pi= 1000$ KHz, $\omega_b/2\pi= 10$ KHz.}\label{fig2}
\end{figure*}
The amplitude of the output field corresponding to the weak probe field can be written as
\begin{equation}
E_{out}=\dfrac{\sqrt{2\kappa}}{\varepsilon_{p}}c_{-}=\mu + i \nu,
\end{equation}

where, the real and imaginary parts, $\mu$ and $\nu$ respectively, account for the inphase and out of phase quadratures of the output field at probe frequency corresponding to the absorption and dispersion. In cavity optomechanics \cite{AKM,Kivshar}, the occurrence of EIT is a result of strong Fano interferences under the simultaneous presence of pump and probe fields which generate a radiation pressure force at the beat frequency $\delta = \omega_p - \omega_l$. The frequency of the pump field $\omega_l$ is shifted to the anti-Stokes frequency $\omega_l + \omega_b$, which is degenerated with the probe field. Destructive interference between the anti-Stokes field and the probe field can suppress the build-up of an intracavity probe field, and result in the narrow transparency window when both the beat frequency $\delta$ and cavity detuning $\Delta$ are resonant with the mechanical frequency $\omega_b$ \cite{Weis,AH,Agarwal,Akram}. Since EIT results from the interference of different frequency contributions, it is well known that we can expect Fano profiles in EIT under certain conditions \cite{Peng,Akram,Qu,GSAgarwal}.

We first present the absorption profiles as a function of normalized detuning $\overline{\delta}/\omega_b$ ($\overline{\delta}=\delta-\omega_b$) in Fig.~\ref{fig2}(a) and (b) at the fixed effective coupling strength $g=0.1\omega_b$ and $g=1\omega_b$, respectively. At this point, starting from $\Delta=0.7\omega_b$, we continuously tuned the frequency $\Delta$ of the high-$Q$ nanocavity, such that it approached to the vicinity of frequency of the Bogoliubov mode $\omega_b$ (see the middle curve). As the frequency-detuning between the modes gradually decreased, the spectral features in the absorption profile exhibit less asymmetry as shown in the lower panels of Fig.~\ref{fig2} (from bottom to top). Consequently, we first observe a series of asymmetric Fano lineshape with the peak located closer to the lower-detuning side for $\Delta=0.7,0.8,0.9 \times\omega_b$, and then a usual transparency window occurs at the exact resonance $\Delta=\omega_b$ (Fig.~\ref{fig2}(a), middle panel). Thus, the asymmetry of Fano resonances decreased as we approached to resonant-detuning $\Delta=\omega_b$, where the phenomenon of EIT takes place, and Fano asymmetry disappears \cite{Peng}. As the detuning $\Delta$ is further increased ($\Delta > \omega_b$), the spectral response of the probe field started to increase again leading to the emergence of Fano lineshapes whose peaks were also located closer to zero-detuning (Fig.~\ref{fig2}(a), upper panels). Figure~\ref{fig2}(b) shows the same curves for $g=1\omega_b$, which depicts the broadening of Fano profiles as well as the EIT window, upon increasing the coupling strength. Thus, more asymmetry can be observed at low values of coupling strength as the resonant region is relatively narrower, and at high frequency-detuning difference ($\Delta \neq \omega_b$) \cite{Peng}. Analytically, for the resonance region $\delta \sim \omega_b$, we obtain the following Fano relation \cite{Ott},
\begin{equation} 
\mu \approx \dfrac{2}{1+q^2}\dfrac{(x+q)^2}{1+x^2}. \label{10}
\end{equation} 
where $x = \frac{\nu + U_{eff}-\omega_b}{\Gamma} -q$, $\Gamma=\frac{2\kappa\Delta g}{\kappa^2+\Omega^2}$, $q =-\Omega/\kappa$, and $\Omega=\Delta-\omega_b$. Interestingly, the absorption profile in Eq.~(\ref{10}) has the same form as the original Fano formula \cite{Fano,Kivshar} with minimum (zero) and maximum at $x = -q$ and $x = 1/q$, respectively. Here, the asymmetry parameter $q$ is related to the frequency offset $\Omega$ which controls the emergence of the asymmetric Fano profiles. Physically it means that the anti-Stokes process is not resonant with the cavity frequency as $\Delta \neq\omega_b$, this condition paves the way towards the observation of tunable Fano resonances \cite{GSAgarwal,Akram,Peng,Li}.
\begin{figure}[!hb]
\includegraphics[width=0.45\textwidth]{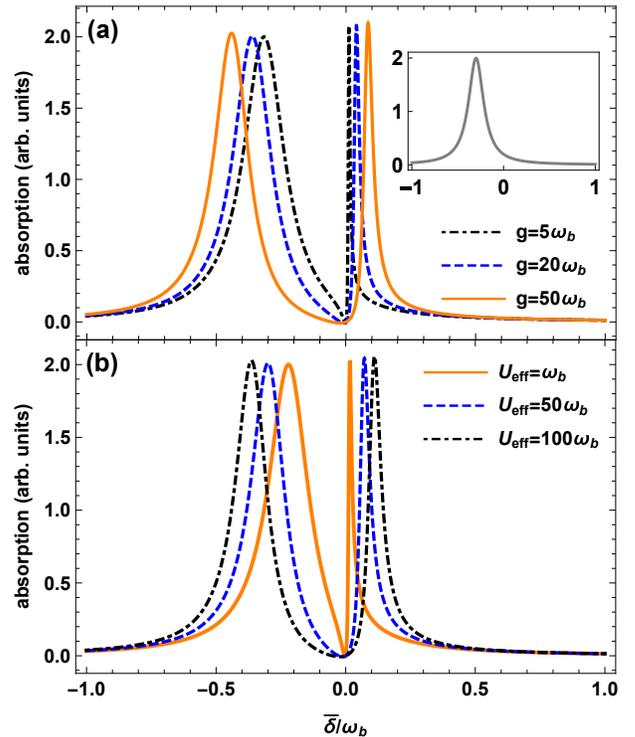}
\caption{(Color online) Fano resonances in the absorption profiles are shown for (a) $g=5\omega_b,20\omega_b,50\omega_b$ corresponding to dot-dashed, dashed and solid curves respectively, and (b) $U_{eff}/\omega_b=1,50,100$, solid, dashed and dot-dashed curves respectively. Here,  $\Delta=0.8\omega_b$, and all the other parameters are the same as in Fig.~\ref{fig2}.}\label{fig3}
\end{figure}
\begin{figure*}[ht]
\includegraphics[width=0.325\textwidth]{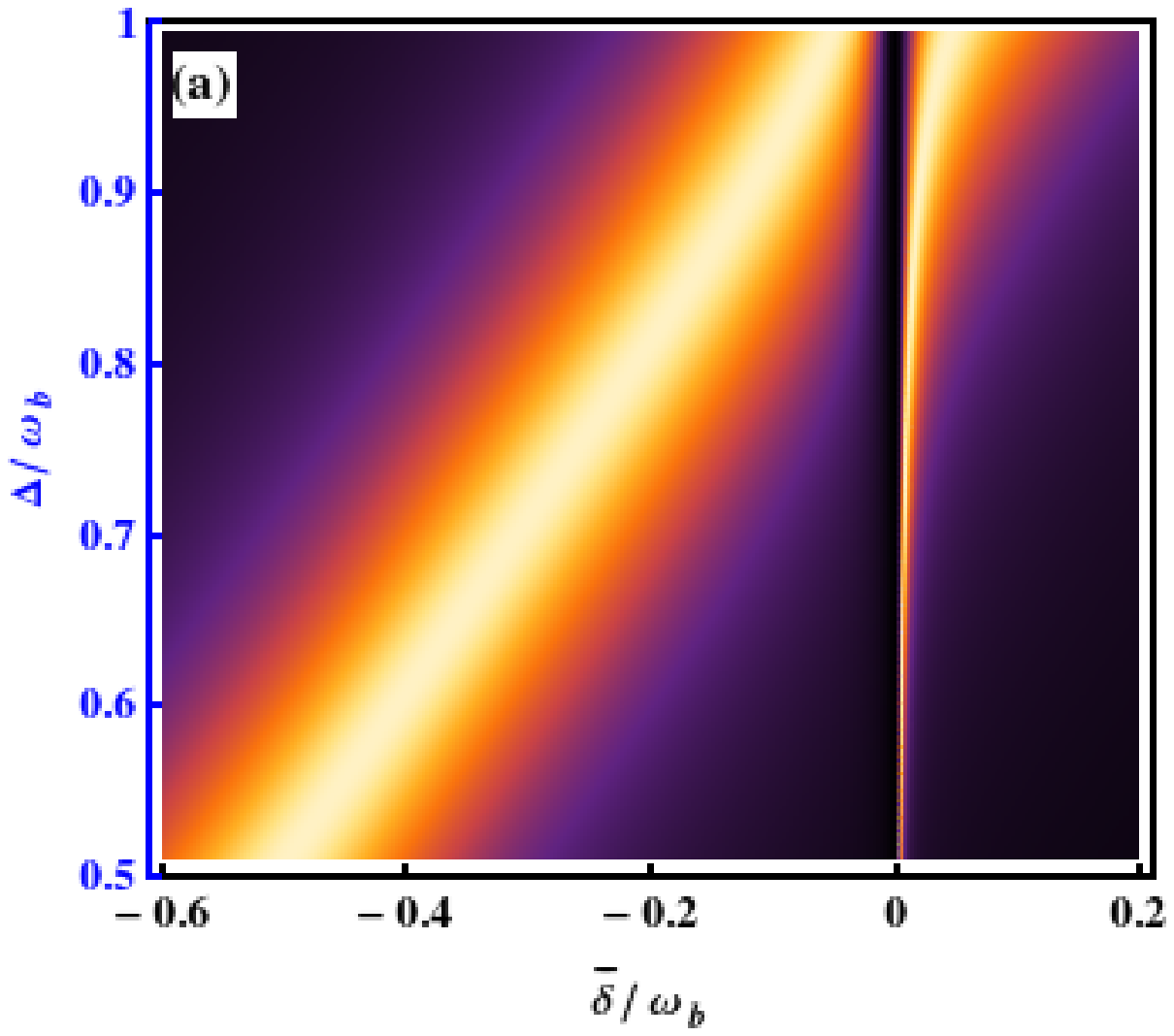}
\includegraphics[width=0.325\textwidth]{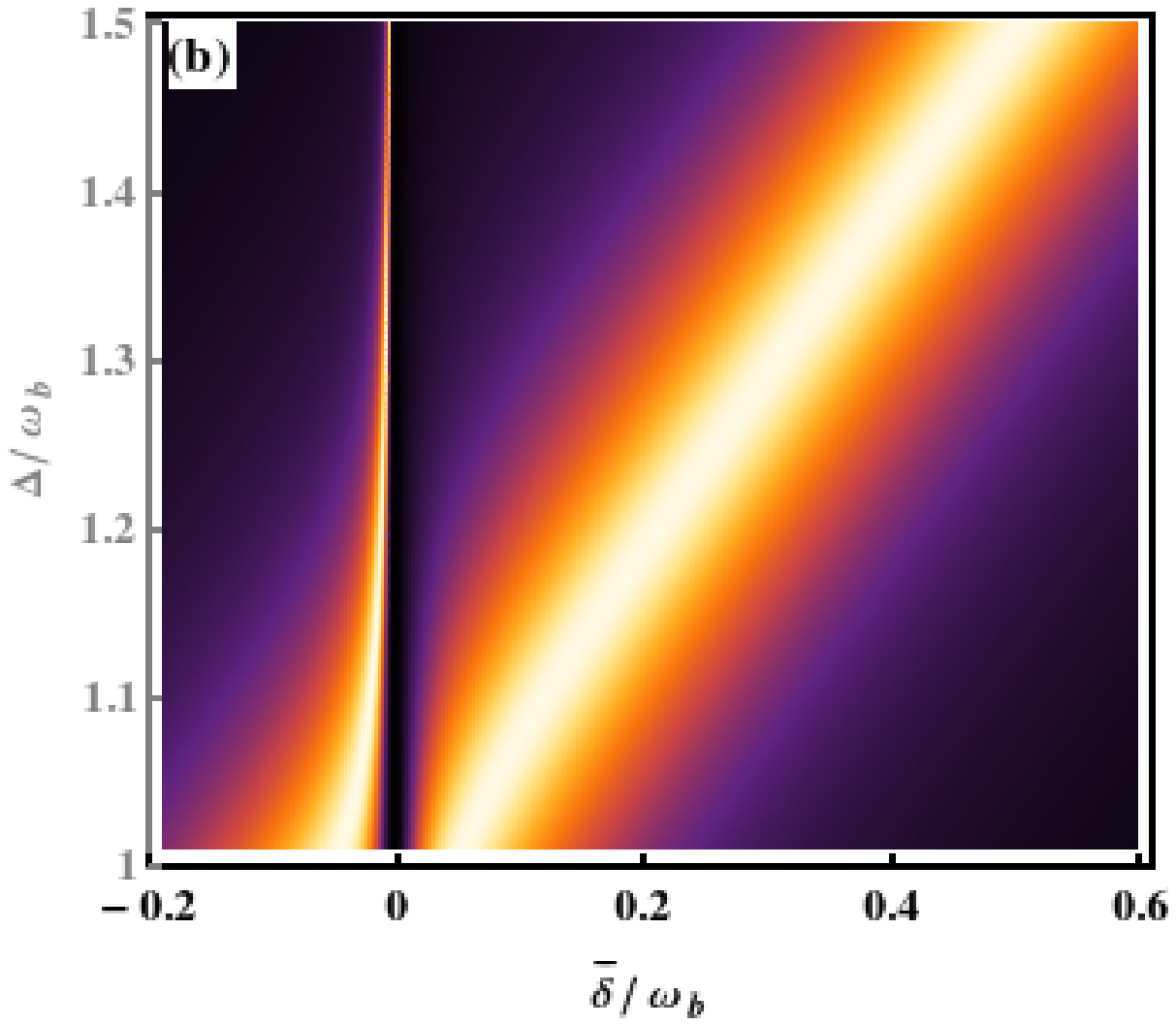}
\includegraphics[width=0.325\textwidth]{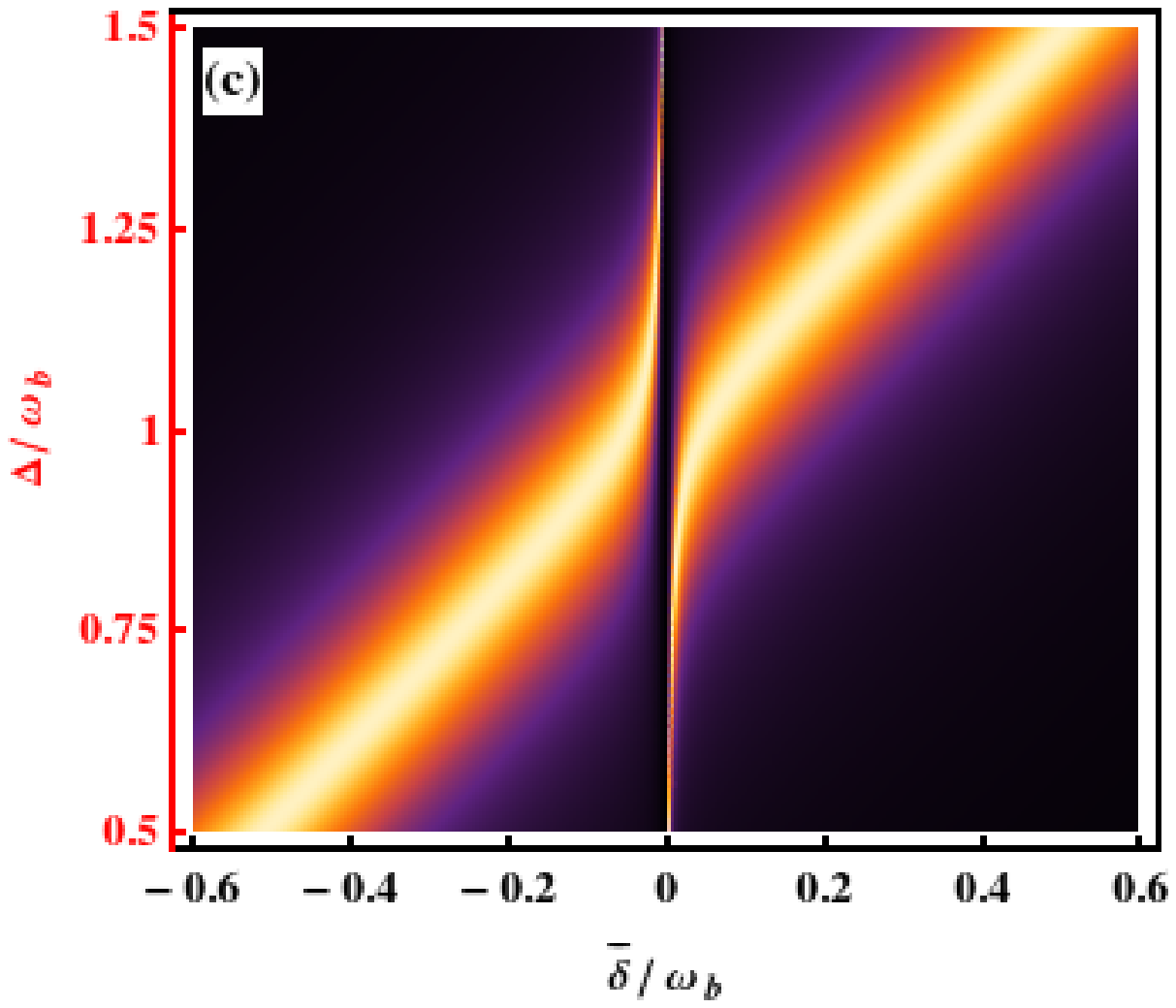}
\caption{(Color online) Spectrum of Fano resonances in the absorption profiles are shown for a range of specific detuning: (a) For $\Delta/\omega_b=0.5-1$ i.e. $\Delta<\omega_{b}$, containing the profiles as shown in Fig.~\ref{fig2}(a). (b) For $\Delta/\omega_b=1.0-1.5$, which shows that narrow and broad regions in Fano profiles can be flipped by suitably tuning the detuning at other side (i.e. $\Delta>\omega_b$) of the resonance $\Delta\sim \omega_b$. (c) For $\Delta/\omega_b=0.5-1.5$, which reflects that Fano spectrum is symmetric around $\Delta \sim \omega_b$. All the other parameters are the same as in Fig.~\ref{fig2}}\label{fig4}
\end{figure*}

Next, we discuss the emergence of Fano resonances in the absorption profile with respect to the coupling strength $g$, and explain how the atom-atom interaction in the BEC can effect the asymmetry of these resonances. We present the absorption profiles as a function of normalized detuning $\overline{\delta}/\omega_b$  in Fig.~\ref{fig3}(a) for different values of the coupling strength $g$. It is noted that, upon increasing the coupling strength, the resonance around $\overline{\delta} \sim \omega_b$ broadens and asymmetry reduces as shown in Fig.~\ref{fig3}(a). Note that, in the absence of the coupling strength $g=0$, the two peaks in Fig.~\ref{fig3}(a) converge to a single peak even when $U_{eff}\neq 0$ as indicated from equation (\ref{eq6}), making a standard Lorentzian absorption peak as seen in the inset (gray curve) in Fig.~\ref{fig3}(a). We examine the effect of finite two-body atom-atom interaction on the Fano profiles in Fig.~\ref{fig3}(b). Note that, in the presence of the finite coupling strength, the narrow Fano profile shown as solid line, goes to the relatively broad resonance as we continuously increase the atom-atom interaction $U_{eff}$, and the asymmetry reduces in the vicinity of $\overline{\delta} \sim \omega_b$. Thus, a high degree of asymmetry in the Fano profiles can be seen for low values of the coupling strength and weak atom-atom interaction. The profiles shown in Fig.~\ref{fig3}(a) and \ref{fig3}(b) have common minimum or zero, thus these profiles are Fano resonances \cite{Ott,GSAgarwal}. It is worthnoting that, as compared to the double-cavity \cite{Qu} and hybrid atom-cavity \cite{Akram} optomechanical systems, the Fano resonances in the present model can be controlled by adjusting the atom-atom interaction, and the fluctuations associated with BEC (explained below).

\enlargethispage{5.5\baselineskip}
To further shed light on the salient features of Fano resonances, we present the Fano spectra for a range of specific detuning as density plots in Fig.~\ref{fig4}(a-c) around both sides of the resonance $\Delta \sim \omega_b$. The Fano profiles obtained in the lower panels of Fig.~\ref{fig2} (for $\Delta < \omega_b$) belongs to the Fano spectrum as shown in Fig.~\ref{fig4}(a). On the other hand, for $\Delta>\omega_b$ while approaching the crossing region \cite{Bird}, we note that the narrow and broad regions in the Fano profiles can be flipped showing the interchange in symmetry around $\Delta \sim \omega_{b}$ as shown in Fig.~\ref{fig4}(b) and upper panel of Fig.~\ref{fig2}. Fig.~\ref{fig4}(c) encapsulates the Fano spectrum for both sides of the resonance $\Delta \sim \omega_b$, i.e. $\Delta/\omega_b \in [0.5,1.5]$. It is clear from this figure, which also contains the Fano profiles shown in Fig.~\ref{fig2}, that the two different sets of Fano spectrum in Fig.~\ref{fig2} (upper and lower panels) connect smoothly to each other. Most notably, the combination of the two spectra allows the full structure of the anti-crossing between broad and narrow regions to become clear, in agreement with previous reports \cite{Bird,Christ}.
\begin{figure}[b]
\includegraphics[width=0.47\textwidth]{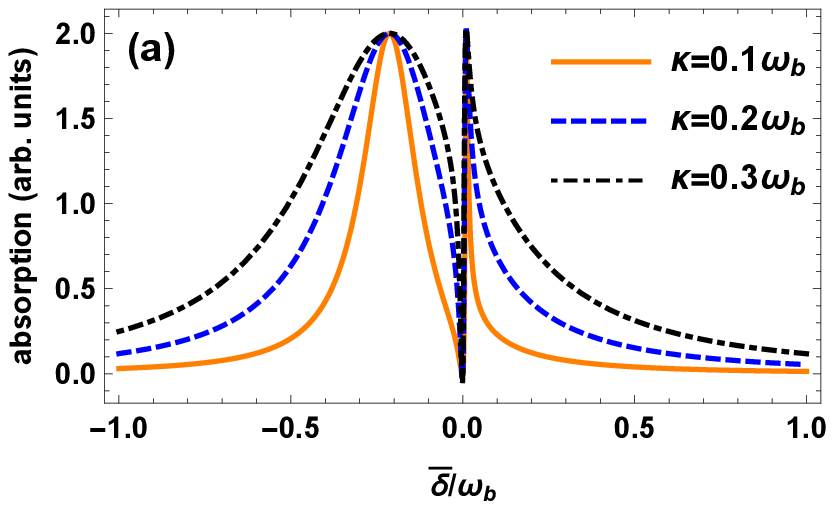}
\includegraphics[width=0.46\textwidth]{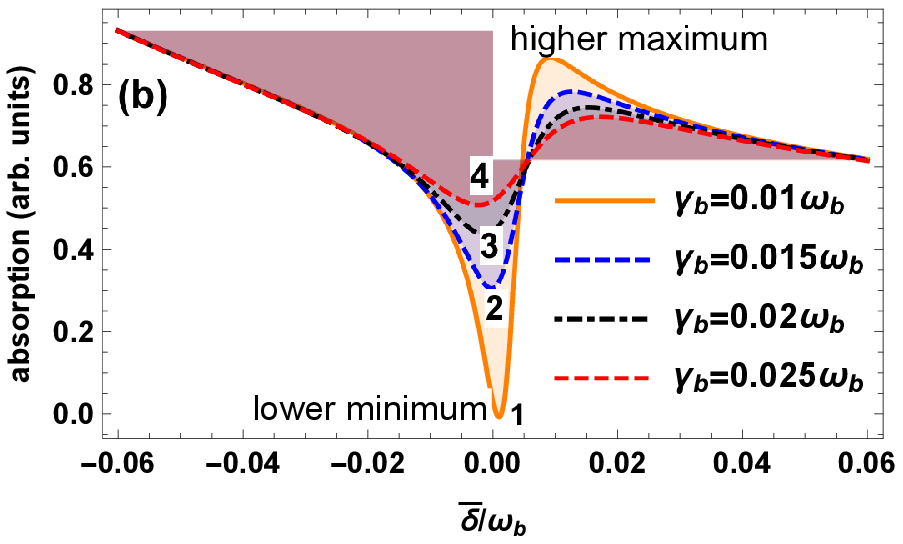}
\caption{(Color online) Fano resonances in the absorption profiles are shown for different values of (a) cavity decay $\kappa$, and (b) condensate (Bogoliubov mode) fluctuations $\gamma_{b}$. In (b), we show BEC-cavity setup can pave the way to observe original Fano-like resonances \cite{Kivshar} with a lower minimum and a higher maximum. All the other parameters are same as in Fig.~\ref{fig2}.}\label{ukfano}
\end{figure}

The existence of Fano line shapes in BEC-OMS can be understood by noting that the response of the Bogoliubov mode (BEC) features the narrow spectral width, whereas the cavity modes have orders of magnitude higher spectral width and therefore, act as continuum channels \cite{Ott}. As discussed earlier, the interference of a narrow bound state with a continuum is known to give rise to asymmetric Fano resonances \cite{Fano,Kivshar,Ott,Peng}. It is worth noting that EIT occurs when the system meet the resonance condition, that is, $\Delta=\omega_b$. However, due to the present non-resonant interactions ($\Delta \neq \omega_b$) the symmetry of the EIT window is transformed into asymmetric Fano shapes. Moreover, for optomechanical systems in general, we have two coherent processes leading to the building up of the cavity field: (i) the direct building up due to the application of strong pump and weak probe field, and (ii) the building up due to the two successive nonlinear frequency conversion processes between optical mode and a mechanical mode \cite{AKM,Qu}. These two paths contribute in the interference which leads to the emergence of Fano profiles in the probe absorption spectrum. Hence, broad and narrow regions can be controlled by appropriately adjusting the system parameters in a nanocavity containing BEC.

Furthermore, we note that the cavity decay rate $\kappa$ and the decay rate of the Bogoliubov mode $\gamma_b$ plays a vital role in the emergence and control of the Fano resonances. In Fig.~\ref{ukfano}(a), we indicate that the Fano resonances are sensitive to the cavity decay rate $\kappa$. On increasing $\kappa$, the resonance around $\overline{\delta} \sim \omega_b$ becomes narrow. These narrow profiles are of primary significance in precision spectroscopy, metrology and high efficiency x-ray detection \cite{Heeg}.

At this point, we emphasize that the Fano profiles as explained above, have a fixed (common) minimum. Nevertheless, in Fig.~\ref{ukfano}(b), we show the magnified Fano resonances for different values of the decay rate of the Bogoliubov mode $\gamma_b$, which remarkably yields different minimum points as pointed out by the numbers $1-4$. Thus, we observe a higher maximum, and correspondingly, a lower minimum in the Fano profiles as shown in Fig.~\ref{ukfano}(b). This effect is analogous to the result in the context of photoionization, in which the value of the minimum depends on the radiative effects \cite{Fano,Kivshar}. Hence, our model not only allows the flexible coherent control to tune the Fano profiles with additional parametric choice (namely the atom-atom interaction and the decay rate of the Bogoliubov mode) unlike previous schemes \cite{Peng,Qu,Akram}, but also paves the way towards the observation of original Fano profiles in a single experimental setup with promising applications, for example, in x-ray detection \cite{Heeg} and ultra-sensitive sensing for biofluid diagnostics \cite{Ameen}.
\section{Slow light in the probe transmission}\label{sec4}
\enlargethispage{6.5\baselineskip}
\begin{figure}[b]
\includegraphics[width=.49\textwidth]{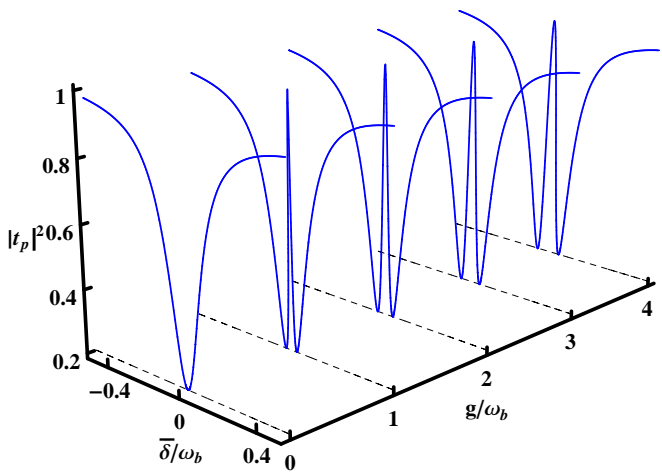}
\caption{(Color online) The transmission $|t_p|^2$ of the probe field as functions of normalized probe detuning $\overline{\delta}/\omega_b$ for $\Delta=\omega_b$ and $g/\omega_b=0,1,2,3,4$, respectively. All the other parameters are the same as in Fig.~\ref{fig2}.}\label{tr1}
\end{figure}
\begin{figure}[t]
\includegraphics[width=.45\textwidth]{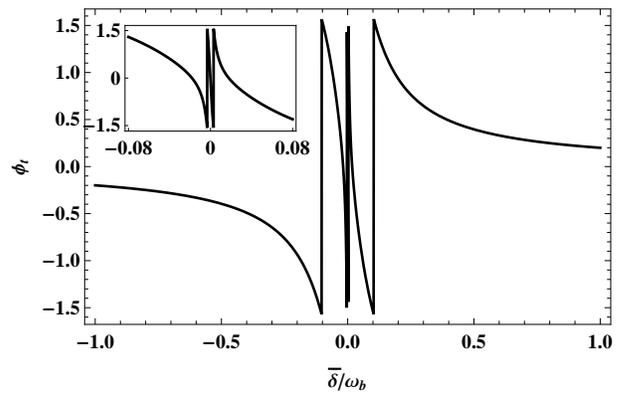}
\caption{(Color online) We plot phase $\phi_t$ of the probe field as function of the normalized probe detuning $\overline{\delta}/\omega_b$ for $\Delta=\omega_b$. The inset shows the rapid change around resonance $\overline{\delta}\sim \omega_b$. All the other parameters are the same as in Fig.~\ref{tr1}.}\label{ph}
\end{figure}

In this section, we analyze the probe field transmission in BEC-cavity setup by utilizing the Bose-Hubbard model. Since both the mirrors are fixed, the interaction between the optical mode and the Bogoliubov mode is analogous to the case of single ended cavity \cite{AH} which yields normal dispersion in probe phase, that allows the slow light propagation in the probe transmission. In Fig.~\ref{tr1}, we show the transmission $|t_p|^2$ of the probe field as a function of normalized probe detuning ($\overline{\delta}/\omega_b$), versus the normalized coupling strength $g/\omega_b$. We observe that usual Lorentzian curve appears in the transmission spectrum for $g=0$ at $\overline{\delta} = 0$ ($\overline{\delta} = \omega_b$) in Fig.~\ref{tr1}. However, the transparency window \cite{Harris,Fleischhauer,Weis,AH,Agarwal} occurs in the probe transmission, once the coupling between the cavity field and the condensate mode is present. Figure.~\ref{tr1} depicts that the transparency window in the transmission spectrum broadens and becomes more prominent by continuously increasing the effective coupling strength $g$. Importantly, unlike traditional optomechanical systems, in addition to the pump laser or the coupling strength, width of the transparency window can effectively be controlled by the atom-atom interaction in our model, as explained in the previous section.
\begin{figure}[hb]
\includegraphics[width=.49\textwidth]{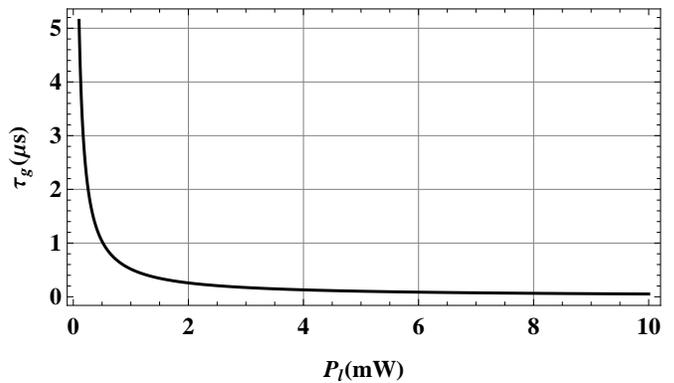}
\caption{(Color online) The group delay $\tau_g$ versus the pump power $P_{l}$ is presented, which shows the pulse delay (subluminal behavior) of the order 5 $\mu s$ in the probe transmission. All the other parameters are the same as in Fig.~\ref{tr1}. }\label{delay}
\end{figure}

In Fig.~\ref{ph}, we plot the phase of the probe field versus the normalized probe detuning for $\Delta=\omega_b$. Due to coupling of the cavity field with BEC, the phase of the probe field undergoes a sharp enhancement in the resonant region $\overline{\delta} \sim \omega_b$, yielding the rapid phase dispersion \cite{Akram2,AH}. This rapid phase dispersion \cite{AH,Akram2} indicates that group delay of the probe field can be changed significantly through the BEC cavity setup. In Fig.~\ref{delay}, we show group delay $\tau_g$ versus the pump power $P_l$. Since the group delay is positive, which reveals that the characteristics of the transmitted probe field exhibits the slow light effect \cite{AH,Akram2,Boyd,Milonni}.

\begin{figure}[t]
\includegraphics[width=.4\textwidth]{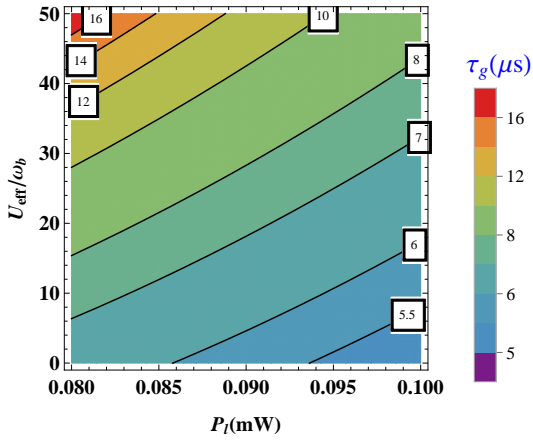}
\caption{(Color online) The group delay $\tau_g$ versus the pump power $P_{l}$ and $U_{eff}$ is presented. The contour plot shows that pulse delay $\tau_g$ (subluminal behaviour) can further be enhanced by increasing the atom-atom interaction. All the other parameters are the same as in Fig.~\ref{tr1}.}\label{del}
\end{figure}
\begin{figure}[b]
\includegraphics[width=.5\textwidth]{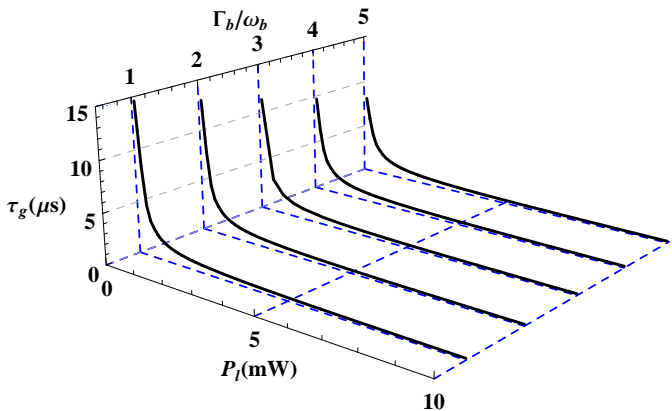}
\caption{(Color online) The group delay $\tau_g$ versus the pump power $P_{l}$ and normalized condensate fluctuations $\Gamma_b/\omega_b$ is presented, which shows that pulse delay (subluminal behaviour) decreases with the increase in the fluctuations of the condensate mode. All the other parameters are the same as in Fig.~\ref{tr1}.}\label{del2}
\end{figure}
\begin{figure}[t]
\includegraphics[width=.45\textwidth]{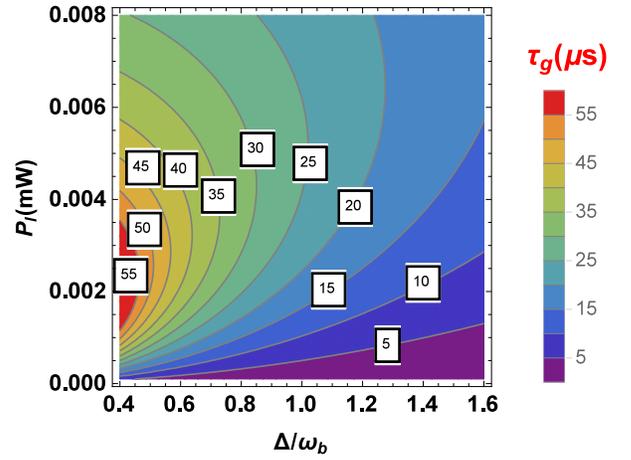}
\caption{(Color online) The group delay $\tau_g$ versus the pump power $P_{l}$ and $\Delta/\omega_b$ is presented, which is related to the Fano spectrum as shown in Fig.~\ref{fig3}(c). The contour plot shows the pulse delay $\tau_g$ (slow light) for both the regimes, $\Delta <\omega_b$ and $\Delta > \omega_b$, where the former regime yields high slow light effect. All the other parameters are the same as in Fig.~\ref{tr1}.}\label{delf}
\end{figure}
The slow light propagation in the presence of coupling between the optical mode and condensate mode is analogous to the slow light propagation as reported for the mirror-field coupling in standard optomechanical cavity setups \cite{AH,Tarhan,Jiang}. However, in the former section, we noted that Fano resonances in the probe absorption spectrum are significantly altered by tuning the atom-atom interaction and also affected by the condensate fluctuations. In order to quantify the effect of atom-atom interaction $U_{eff}$ and the fluctuations of the condensate mode $\gamma_{b}$ on the probe transmission, we present the group delay $\tau_{g}$ in Fig.~\ref{del} and Fig.~\ref{del2}. Interestingly from Fig.~\ref{del}, we see that on increasing the atomic two-body interaction ($U_{eff}$), the magnitude of group delay increases. Hence, slow light effect becomes more prominent with large atom-atom interaction, uniquely exist in BEC-cavity setup, and therefore, reflects a clear advantage for the realization of optical memory \cite{Simon,Phillips} over previous optomechanics reports on slow light \cite{AH,Akram2,Tarhan,Zhu,Zhan,Jiang}.

Furthermore, the effect of the condensate fluctuations $\gamma_{b}$ on the group delay can be understood from Fig.~\ref{del2}. We note that by tuning $\gamma_b$, the group delay remains unaffected or robust (not shown) until condensate fluctuations attains higher values. However, on increasing $\Gamma_{b}$, the magnitude of group delay decreases gradually (where, $\Gamma_{b}/2\pi =4.1$ KHz \cite{Esslinger} $\approx \omega_b \times \gamma_b$ in magnitude  i.e. $\Gamma_{b}>>\gamma_{b}$), which indicates that slow light effect in our scheme is capable for optical storage applications \cite{Simon,Jobez} owning to the great flexibility against the condensate fluctuations. More importantly, unlike single-ended optomechanical systems \cite{AH} and hybrid atom-cavity systems \cite{Akram2}, the delay-bandwidth product in the present case can be greatly enhanced by continuously increasing atom-atom interaction, and therefore serve as a way to store an optical pulse \cite{Simon,Jobez}. 

Finally, we remark that the above discussion of slow light through the probe transmission is explained under the EIT effect, i.e. for $\Delta=\omega_b$. In Fig.~\ref{delf}, we show the group delay versus pump power for $\Delta/\omega_b =0.5-1.5$. Note that, this parametric regime is directly associated to the Fano spectrum given in Fig.~\ref{fig3}(c). We see that for low one photon detuning i.e. $\Delta < \omega_b$, slow light effect is enormous as group delay of the order $\tau_g=55~\mu s$ is observed around $\Delta\approx 0.4 \omega_b$. As the detuning frequency difference is continuously increased, the magnitude of slow light goes down to $\tau_g=5~\mu s$ for $\Delta\approx 1.6 \omega_b$. Moreover, we remark that the group delay $\tau_g>0$ in Figs.~(8-11), which indicates the slow light behavior of the transmitted probe beam, accounts for the case of normal dispersion which occurs very close to the resonant region in Fig. \ref{ph}. However, for the case of anomalous dispersion, one would always obtain $\tau_g < 0$, which then represents the superluminal or fast light behavior of the transmitted probe beam (see e.g. Refs.~\cite{AH,Akram2,Milonni}).

Presently, a plethora of suggestions have been made to realise an array of connected (coupled) cavities with a moving-end mirror, for example Refs.~\cite{Chiara,Kimble2,Xuereb}. Moreover, our BEC-cavity optomechanics setup has also been realized to transfer the quantum state of light fields to the collective density excitations of BEC \cite{Bhattacherjee2}. For the reason, the present scheme may serve as a building block to realize efficient optical delay lines for quantum networks \cite{AH,Jobez}.
\section{Conclusions}\label{sec5}
In conclusion, we have studied Fano resonances, slow light and its enhancement in a nano cavity using Bose-Einstein Condensate. It is shown that in the simultaneous presence of pump and probe fields, the condensate coherently couples with the cavity standing wave field to exhibit quantum destructive interference. As a result, a typical Fano resonance, which can also be controlled through the strength of atom-atom interaction, is emerged. Moreover, the anti-crossing of Fano spectrum and the existence of original Fano profiles in the presence of BEC have been noted. Furthermore, in addition to EIT, the occurrence of slow light in the probe transmission is studied, and the group delay is analyzed for the atomic two-body interaction and fluctuations of the condensate. Unlike previous schemes \cite{AH,Akram2,Tarhan,Zhu,Zhan,Jiang}, by utilizing the Bose-Hubbard model in the present setup, we show that the slow light effect can further be enhanced by increasing the atom-atom interaction.

Despite the promising applications for ultra-sensitive sensing \cite{Ameen,Kivshar,Heeg} via Fano resonances, interesting continuation of this work would be to extend the slow light experiment using BEC by incorporating the extended Bose-Hubbard model in optomechanical systems \cite{Chiara,Rossini,Miao,Corrielli,Zoller}. For instance, the extended BH model can lead to the large group delays due to additional interference by encompassing long-range atom-atom interactions within BEC \cite{Zoller}. Thus, promising optical information storage applications, such as reported in Refs.~\cite{Simon,Jobez,Phillips} for the heralded long-distance quantum communication \cite{Kimble}, are expected.
\section*{Acknowledgments}
M.J.A. and F.S. acknowledge Higher Education Commission (HEC), Pakistan, for financial support through Grant No. \#HEC/20-1374.  F.G. acknowledges HEC for financial support through Grant No. \#HEC/NRPU/20-2475.

\end{document}